\documentstyle[12pt]{article}

\font\twlgot =eufm10 scaled \magstep1 \font\egtgot =eufm8
\font\sevgot =eufm7 \font\twlmsb =msbm10 scaled \magstep1
\font\egtmsb =msbm8 \font\sevmsb =msbm7

\newfam\gotfam

\textfont\gotfam\twlgot \scriptfont\gotfam\egtgot
\scriptscriptfont\gotfam\sevgot

\newfam\msbfam
\textfont\msbfam\twlmsb \scriptfont\msbfam\egtmsb
\scriptscriptfont\msbfam\sevmsb

\def\pBbb{\relax\ifmmode\expandafter\Bb\else\typeout{You cann't use
Bbb in text mode}\fi}
\def\Bb #1{{\fam\msbfam\relax#1}}

\def\thebibliography#1{\section*{References}\list
  {[\arabic{enumi}]}{\settowidth\labelwidth{#1}\leftmargin\labelwidth
    \advance\leftmargin\labelsep
    \usecounter{enumi}}
    \def\newblock{\hskip .11em plus .33em minus .07em}
    \sloppy\clubpenalty4000\widowpenalty4000
    \sfcode`\.=1000\relax}

\newcommand{\beq}{\begin{equation}}
\newcommand{\eeq}{\end{equation}}
\newcommand{\ben}{\begin{eqnarray}}
\newcommand{\een}{\end{eqnarray}}
\newcommand{\be}{\begin{eqnarray*}}
\newcommand{\ee}{\end{eqnarray*}}
\newcommand{\bea}{\begin{eqalph}}
\newcommand{\eea}{\end{eqalph}}

\newcounter{eqalph}
\newcounter{equationa}
\newcounter{remark}
\newcounter{example}
\newcounter{theorem}
\newcounter{proposition}
\newcounter{lemma}
\newcounter{corollary}
\newcounter{definition}
\def\theremark{\arabic{remark}}
\def\thetheorem{\arabic{theorem}}

\def\thedefinition{\arabic{definition}}

\newenvironment{eqalph}{\stepcounter{equation}
\setcounter{equationa}{\value{equation}} \setcounter{equation}{0}

\begin{eqnarray}}{\end{eqnarray}
\setcounter{equation}{\value{equationa}}}

\newcommand{\mar}[1]{}

\begin{document}
\hbox{}

{\parindent=0pt

{\large \bf On incompleteness of classical field theory}
\bigskip

{\sc G. Sardanashvily}

{\sl Department of Theoretical Physics, Moscow State University,
117234 Moscow, Russia}

\bigskip
\bigskip

\begin{small}

Classical field theory is adequately formulated as Lagrangian
theory on fibre bundles and graded manifolds. One however observes
that non-trivial higher stage Noether identities and gauge
symmetries of a generic reducible degenerate Lagrangian field
theory fail to be defined. Therefore, such a field theory can not
be quantized.

\end{small}

 }

\bigskip
\bigskip
\bigskip

Contemporary quantum field theory (QFT) is mainly developed as
quantization of classical fields.  In contrast with QFT, classical
field theory can be formulated in a strict mathematical way
\cite{ijgmmp08,book09}.

Observable classical fields are an electromagnetic field, Dirac
spinor fields and a gravitational field on a world real smooth
manifold. Their dynamic equations are Euler--Lagrange equations
derived from a Lagrangian. Classical non-Abelian gauge fields and
Higgs fields also are considered. Basing on these models, one
studies Lagrangian theory of classical fields on an arbitrary
smooth manifold $X$ in a very general setting. Geometry of
principal bundles is known to provide the adequate mathematical
formulation of classical gauge theory. Generalizing this
formulation, one defines even classical fields as sections of
smooth fibre bundles and, accordingly, develop their Lagrangian
theory as Lagrangian theory on fibre bundles.

Note that, treating classical field theory, we are in the category
of finite-dimensional smooth real manifolds, which are Hausdorff
second-countable and paracompact. Let $X$ be such a manifold. If
classical fields form a projective $C^\infty(X)$-module of finite
rank, their representation by sections of a fibre bundle follows
from the well-known Serre--Swan theorem.

Lagrangian theory on fibre bundles is adequately formulated in
algebraic terms of the variational bicomplex of exterior forms on
jet manifolds \cite{and,cmp04,book09,tak2}. This formulation is
straightforwardly extended to Lagrangian theory of even and odd
fields by means of the Grassmann-graded variational bicomplex
\cite{barn,lmp08,cmp04,book09}. Cohomology of this bicomplex
\cite{book09,ijgmmp07} provides the global first variational
formula for Lagrangians and Euler--Lagrange operators, the first
Noether theorem and conservation laws in a general case of
supersymmetries depending on derivatives of fields of any order.

Note that there are different descriptions of odd fields on graded
manifolds. Both graded manifolds and supermanifolds are described
in terms of sheaves of graded commutative algebras
\cite{bart,book00}. However, graded manifolds are characterized by
sheaves on smooth manifolds, while supermanifolds are constructed
by gluing of sheaves on supervector spaces. Treating odd fields on
a smooth manifold $X$, we follow the Serre--Swan theorem
generalized to graded manifolds \cite{jmp05a,book09}. It states
that, if a Grassmann $C^\infty(X)$-algebra is an exterior algebra
of some projective $C^\infty(X)$-module of finite rank, it is
isomorphic to the algebra of graded functions on a graded manifold
whose body is $X$.

Quantization of Lagrangian field theory essentially depends on its
degeneracy characterized by a family of non-trivial reducible
Noether identities \cite{barn,lmp08,gom}. A problem is that any
Euler--Lagrange operator satisfies Noether identities which
therefore must be separated into the trivial and non-trivial ones.
In accordance with general theory of Noether identities of
differential operators \cite{book09,oper} Noether identities of
Lagrangian theory are represented by cycles of a certain chain
complex, whose boundaries are treated as trivial Noether
identities and whose homology describes non-trivial Noether
identities modulo the trivial ones \cite{jmp05a,lmp08,book09}.
Lagrangian field theory is called degenerate if its
Euler--Lagrange operator satisfies non-trivial Noether identities.
These Noether identities obey first-stage Noether identities,
which in turn are subject to the second-stage ones, and so on.
Higher-stage Noether identities must also be separated into the
trivial and non-trivial ones. To describe non-trivial
$(k+1)$-stage Noether identities, one must assume the following.

(i) Non-trivial $k$-stage Noether identities are generated by a
projective $C^\infty(X)$-module of finite rank. In this case,
$(k+1)$-stage Noether identities are represented by $(k+2)$-cycles
of some chain complex.

(ii) This chain complex obeys a certain homology condition. Then
trivial $(k+1)$-stage Noether identities are identified with its
$(k+2)$-boundaries of this complex, and its $(k+2)$-homology
describes non-trivial $(k+1)$-stage Noether identities.

{\it A problem is that degenerate Lagrangian field theory need not
satisfy these conditions, and its non-trivial higher stage Noether
identities fail to be defined in general.}

Degenerate Lagrangian field theory is called reducible if there
exist non-trivial higher stage Noether identities. The hierarchy
of its Noether identities is described by the exact Koszul--Tate
chain complex of antifields possessing the boundary operator whose
nilpotentness is equivalent to all non-trivial Noether and
higher-stage Noether identities \cite{jmp05a,lmp08,book09}.

The inverse second Noether theorem formulated in homology terms
associates to this Koszul--Tate complex the cochain sequence of
ghosts with the ascent operator, called the gauge operator, whose
components are non-trivial gauge and higher-stage gauge symmetries
of Lagrangian field theory \cite{lmp08,jmp09,book09}. It should be
emphasized that the gauge operator unlike the Koszul--Tate one is
not nilpotent, unless non-trivial gauge symmetries are abelian.
This is the cause why an intrinsic definition of non-trivial gauge
and higher-stage gauge symmetries meets difficulties. Defined by
means of the inverse second Noether theorem, non-trivial gauge and
higher-stage gauge symmetries are parameterized by odd and even
ghosts so that $k$-stage gauge symmetry acts on $(k-1)$-stage
ghosts.

{\it Since non-trivial higher stage gauge symmetries are derived
from non-trivial higher stage Noether identities by means of the
inverse second Noether theorem, it may happen that they are not
defined in a general case of degenerate Lagrangian field theory.

Thus one concludes that classical field theory is incomplete
because the degeneracy of Lagrangian field theory fails to be
analyzed in general.}

Gauge symmetries are said to be algebraically closed if this gauge
operator admits a nilpotent extension where $k$-stage gauge
symmetries are extended to $k$-stage BRST transformations acting
both on $(k-1)$-stage and $k$-stage ghosts
\cite{lmp08,jmp09,book09}. This nilpotent extension  is called the
BRST operator. If the BRST operator exists, the cochain sequence
of ghosts is brought into the BRST complex.

The Koszul--Tate and BRST complexes provide the BRST extension of
original Lagrangian field theory by means of antifields and ghosts
which form projective $C^\infty(X)$-modules of finite rank
isomorphic to $C^\infty(X)$-modules of non-trivial Noether
identities and gauge symmetries in accordance with the Serre--Swan
theorem \cite{book09}. This BRST extension is a first step towards
quantization of degenerate Lagrangian field theory in terms of
functional integrals \cite{barn,gom}.

{\it However, degenerate Lagrangian field theory can not be
quantized if its non-trivial Noether identities and gauge
symmetries are not defined. It follows that quantization of
classical fields fails to be a universal principle of constructing
quantum field theory.}

\end{document}